\documentclass{ACFD}
\usepackage{amsmath,amssymb,amsfonts}
\usepackage{graphicx}
\usepackage{booktabs}
\usepackage{listings}
\usepackage{enumitem}
\usepackage{subfig}
\usepackage[ruled,vlined]{algorithm2e}
\usepackage{float}
\usepackage{multirow}
\newcommand\scalemath[2]{\scalebox{#1}{\mbox{\ensuremath{\displaystyle #2}}}}

\definecolor{alizarin}{rgb}{0.82, 0.1, 0.26}
\definecolor{rufous}{rgb}{0.66, 0.11, 0.03}
\definecolor{darkred}{rgb}{0.55, 0.0, 0.0}
\definecolor{bluep}{rgb}{0.2, 0.2, 0.6}
\definecolor{fgreen}{rgb}{0.0, 0.27, 0.13}
\definecolor{maroon}{rgb}{0.76, 0.13, 0.28}
\definecolor{blue(munsell)}{rgb}{0.0, 0.5, 0.69}
\definecolor{codegreen}{rgb}{0,0.6,0}
\definecolor{codegray}{rgb}{0.5,0.5,0.5}
\definecolor{codepurple}{rgb}{0.58,0,0.82}
\definecolor{backcolour}{rgb}{0.95,0.95,0.92}
\definecolor{cadmiumgreen}{rgb}{0.0, 0.42, 0.24}
\definecolor{darkorange}{rgb}{1.0, 0.55, 0.0}
\definecolor{aliceblue}{rgb}{0.94, 0.97, 1.0}
\definecolor{anti-flashwhite}{rgb}{0.95, 0.95, 0.96}
\definecolor{ghostwhite}{rgb}{0.97, 0.97, 1.0}
\definecolor{bleudefrance}{rgb}{0.19, 0.55, 0.91}

\lstdefinelanguage{Regent}
{morekeywords=[1]{import, require, regentlib.start, var, __demand, __cuda, __inline, __index_launch, __vectorize, fill, reads, writes, reduces, colors},
morekeywords=[2]{task, return, region, fspace, ispace, partition, image, preimage, equal, disjoint, aliased, int, double, int1d, int2d, int3d, for, while, do, break, if, end, else, then, elseif, where},
morekeywords=[3]{local},
sensitive=true,
morecomment=[l]{--},
morecomment=[s]{--[[}{--]]},
morestring=[b]",
}

\lstdefinelanguage{CUDA C}
{morekeywords=[1]{int, if, return, __global__, grid, threadsPerBlock, blocksPerGrid},
morekeywords=[2]{void, __global__, threadIdx, blockIdx, blockDim, cudaDeviceSynchronize},
morekeywords=[3]{local},
sensitive=true,
morecomment=[l]{//},
morecomment=[s]{--[[}{--]]},
morestring=[b]",
}

\lstdefinestyle{my_style}{
    breakatwhitespace=false,
    basicstyle=\ttfamily\footnotesize,
    keywordstyle=[1]\bfseries\color{bleudefrance},
    keywordstyle=[2]\bfseries\color{alizarin},
    keywordstyle=[3]\color{darkorange},
    stringstyle = \color{purple},
    backgroundcolor=\color{white},
    breaklines=true,                 
    captionpos=b,                    
    keepspaces=true,                 
    showspaces=false,                
    numbers=left,                    
    numbersep=5pt,
    xleftmargin=10pt,
    showstringspaces=false,
    showtabs=false,                  
    tabsize=2
}

\begin{document}
\title{Regent based parallel meshfree LSKUM solver for heterogenous HPC platforms}
\author{Sanath Salil$^{1*}$, Nischay Ram Mamidi$^2$, Anil Nemili$^{3}$, Elliott Slaughter$^{4}$}
\affiliation{
		$^1$ Department of Computer Science, BITS Pilani-Hyderabad Campus, Hyderabad, India\\
 		\vspace{7pt}
        $^2$ Department of Computer Science, Rutgers University, Camden, USA \\
 		\vspace{7pt}
        $^3$ Department of Mathematics, BITS Pilani-Hyderabad Campus, Hyderabad, India\\
		  \vspace{7pt}
        $^4$Computer Science Research Department, SLAC National Accelerator Laboratory, Stanford University, Menlo Park, USA \\  
  	    \vspace{7pt}  
		$^*$ 
        f20180812@hyderabad.bits-pilani.ac.in
		}
\maketitle
%
\begin{abstract}
{
Regent is an implicitly parallel programming language that allows the development of a single codebase for heterogeneous platforms targeting CPUs and GPUs. This paper presents the development of a parallel meshfree solver in Regent for two-dimensional inviscid compressible flows. The meshfree solver is based on the least squares kinetic upwind method. Example codes are presented to show the difference between the Regent and CUDA-C implementations of the meshfree solver on a GPU node. For CPU parallel computations, details are presented on how the data communication and synchronisation are handled by Regent and Fortran+MPI codes. The Regent solver is verified by applying it to the standard test cases for inviscid flows. Benchmark simulations are performed on coarse to very fine point distributions to assess the solver's performance. The computational efficiency of the Regent solver on an A100 GPU is compared with an equivalent meshfree solver written in CUDA-C. The codes are then profiled to investigate the differences in their performance. The performance of the Regent solver on CPU cores is compared with an equivalent explicitly parallel Fortran meshfree solver based on MPI. Scalability results are shown to offer insights into performance. \\ \\
%

 } 
\end{abstract}
%
\section{Introduction}
It is well known that the accurate computation of fluid flows involving fine grids is computationally expensive. Typically, the computational fluid dynamics (CFD) codes employed for such applications are CPU or GPU parallel. Since modern high performance computing (HPC) platforms are incresanaasingly becoming heterogeneous, the codes written in programming languages such as Fortran, C, Python, and Julia may not exploit the current architecture. On the other hand, developing and maintaining both CPU and GPU versions of the parallel code is tedious and may require more human resources. Note that separate codes have to be written to target NVIDIA and AMD GPUs. Therefore, it is desirable to have a CFD code in a language that can target a heterogeneous platform consisting of any CPU or GPU configuration. Furthermore, it will benefit immensely if the language supports implicit parallelism. The programming language Regent \cite{Regent15} precisely addresses these requirements.    \\ \\ 
Regent is a task-based programming language built on the Legion \cite{6468504} framework that can optimally utilise modern high-performance computing platforms. Programs in Regent consist of tasks written in sequential order. Note that a task can be thought of as a subroutine in Fortran. Each task has specific privileges to operate on sets of data. Regent can infer data dependencies between tasks. Using this information, the Legion runtime can automatically schedule tasks and transfer data within the computing platform while preserving the sequential semantics of the program. Furthermore, applications written in Regent can be executed on various system configurations, including a single CPU, GPU, or a system with multiple nodes consisting of CPUs and GPUs, with virtually no changes. \\ \\
To the best of our knowledge, a rigorous investigation and comparison of the performance of a CFD code in Regent on CPUs and GPUs with the codes written in traditional languages has not been explored. In this research, an attempt has been made to develop a Regent-based meshfree solver for heterogeneous HPC architectures. Here, the meshfree solver is based on the least squares kinetic upwind method (LSKUM) for two-dimensional inviscid flows \cite{lskum-ghosh-aiaa,lskum-ghosh-journal}. 
The computational efficiency of the Regent code is compared with an equivalent CUDA-C meshfree code on a GPU \cite{hipc-paper} and MPI parallel Fortran code on CPUs \cite{regent-paper-lskum}. \\ \\ 
This paper is organised as follows. Section \ref{sec-lskum} presents the basic theory of the meshfree scheme based on LSKUM. Section \ref{sec-regent-lskum-solver}, presents the development of an LSKUM solver based on Regent for heterogeneous architecture. Example codes are shown to show the difference between the Regent and CUDA-C implementations of the solver on a GPU node. For CPU parallel computations, details are presented on how the data communication and synchronisation are handled by Regent code and Fortran+MPI code. To verify the Regent meshfree solver, Section \ref{sec-code-verification} presents the numerical results on the standard inviscid flow test cases for the NACA 0012 airfoil. A detailed analysis of the performance of the Regent meshfree solver on a GPU and CPUs is presented in Sections \ref{sec-per-analysis-gpu} and \ref{sec-per-analysis-cpus}, respectively. Finally, Section \ref{sec-conclusions} presents the conclusions and a plan for future work.  
\section{Theory of Meshfree LSKUM }\label{sec-lskum}
The Least Squares Kinetic Upwind Method (LSKUM) \cite{lskum-ghosh-aiaa,lskum-ghosh-journal} is a kinetic meshfree scheme for the numerical solution of Euler or Navier-Stokes equations. It operates on a distribution of points, known as point cloud. The cloud of points can be obtained from structured, unstructured, or even chimera grids. LSKUM is based on the moment method strategy \cite{kfvs}, where an upwind meshfree scheme is first developed for the Boltzmann equation. Later, suitable moments are taken to get the upwind meshfree scheme for the governing fluid flow equations. We now present the basic theory of LSKUM for the numerical solution of Euler equations that govern the inciscid fluid flows. In two-dimensions ($2D$), the Euler equations are given by
\begin{equation}
\frac{\partial \boldsymbol{U}}{\partial t} + \frac{\partial \boldsymbol{Gx}}{\partial x} + \frac{\partial \boldsymbol{Gy}}{\partial y}  = 0 
\label{ee-conservation-laws}
\end{equation}
Here, $\boldsymbol{U}$ is the conserved vector, $\boldsymbol{Gx}$ and $\boldsymbol{Gy}$ are the flux vectors along the coordinates $x$ and $y$, respectively. These equations can be obtained by taking moments of the $2D$ Boltzmann equation in the Euler limit \cite{kfvs}. In the inner product form, this relation can be expressed as
\begin{equation}
\frac{\partial \boldsymbol{U}}{\partial t} + \frac{\partial \boldsymbol{Gx}}{\partial x} + \frac{\partial \boldsymbol{Gy}}{\partial y} = \left< \boldsymbol{\Psi}, \frac{\partial F}{\partial t} + v_1 \frac{\partial F}{\partial x} + v_2 \frac{\partial F}{\partial y} \right> = 0
\label{ee-be-me}
\end{equation}
Here, $F$ is the Maxwellian velocity distribution function and $\boldsymbol{\Psi}$ is the moment function vector \cite{kfvs}. $v_1$ and $v_2$ are the molecular velocities along the coordinates $x$ and $y$, respectively. Using the Courant-Issacson-Rees (CIR) splitting \cite{cir-splitting} of molecular velocities, an upwind scheme for the $2D$ Boltzmann equation can be constructed as 
\begin{equation}
\frac{\partial F}{\partial t } +  \frac{v_1+\left| v_1 \right| }{2} \frac{\partial F}{\partial x } + \frac{v_1 - \left| v_1 \right| }{2}  \frac{\partial F}{\partial x } +  \frac{v_2+\left| v_2 \right| }{2}   \frac{\partial F}{\partial y } + \frac{v_2 + \left| v_2 \right| }{2}  \frac{\partial F}{\partial y } = 0 
\label{kfvs-2d-be}
\end{equation}
A robust method of obtaining second-order accurate approximations for the spatial derivatives in eq. (\ref{kfvs-2d-be}) is by employing the defect correction procedure \cite{lskum-ghosh-journal}. To derive the desired formulae for $F_x$ and $F_y$ at a point $P_0$, consider the Taylor series expansion of $F$ up to quadratic terms at a point $P_i \in  N\left(P_0\right)$,
\begin{equation}
\Delta F_i =  \Delta x_i F_{x_0} + \Delta y_i F_{y_0} 
+  \frac{\Delta x_i}{2} \left( \Delta x_i F_{xx_0} + \Delta y_i F_{xy_0}  \right)  
 +  \frac{\Delta y_i}{2} \left( \Delta x_i F_{xy_0} 
 +  \Delta y_i F_{yy_0}  \right) + 
  O \left( \Delta x_i, \Delta y_i \right)^3, \; 
 i= 1, \hdots, n
\label{taylor-series-exp-F-second-order-rearranged}
\end{equation}
where $\Delta x_i = x_i - x_0$, $\Delta y_i = y_i - y_0$, $\Delta F_i = F_i - F_0$. $N\left(P_0\right)$ is the set of neighbours or the stencil of $P_0$. Here, $n$ denotes the number of neighbours of the point $P_0$. To eliminate the second-order derivative terms in the above equation, consider the Taylor series expansions of $F_x$ and $F_y$ to linear terms at a point $P_i$
\begin{equation}
\begin{split}
\Delta F_{x_i} = &  F_{x_i} - F_{x_0} = \Delta x_i F_{xx_0} + \Delta y_i F_{xy_0} + O\left( \Delta x_i, \Delta y_i  \right)^2  \\
\Delta F_{y_i} = & F_{y_i} - F_{y_0} = \Delta x_i F_{xy_0} + \Delta y_i F_{yy_0} + O\left( \Delta x_i, \Delta y_i \right)^2 
\end{split}
\end{equation}
Substituting the above expressions in eq.~(\ref{taylor-series-exp-F-second-order-rearranged}), we obtain
\begin{equation}
\Delta F_i =  \Delta x_i F_{x_0} + \Delta y_i F_{y_0} +  \frac{1}{2} \left\lbrace \Delta x_i  \Delta F_{x_i}   +   \Delta y_i \Delta F_{y_i} \right\rbrace +  O \left( \Delta x_i, \Delta y_i \right)^3, \; \; i  = 1, \hdots, n 
\label{taylor-series-second-order-rearranged}
\end{equation}
Define a perturbation in modified Maxwellians, $\Delta \widetilde{F}_i$ as
\begin{equation}
\Delta \widetilde{F}_i  =  \widetilde{  F}_i - \widetilde{F}_0  = \Delta F_i - \frac{1}{2} \left( \Delta x_i \Delta F_{x_i} + \Delta y_i \Delta F_{y_i}  \right)  
\label{modified-perturb-maxwellian}
\end{equation}
Using $\Delta \widetilde{ F}_i $, eq. (\ref{taylor-series-second-order-rearranged}) reduces to
\begin{equation}
\Delta \widetilde{ F}_i   =  \Delta x_i F_{x_0} +  \Delta y_i F_{y_0}  +  O \left( \Delta x_i, \Delta y_i \right)^3, \; i  = 1, \hdots, n 
\label{taylor-series-exp-F-second-order-modified-final}
\end{equation}
For $n \geq 3$, eq. (\ref{taylor-series-exp-F-second-order-modified-final}) leads to an over-determined linear system of equations. Using the least-squares principle, the second-order approximations to $F_x$ and $F_y$ at the point $P_0$ are given by
\begin{equation}
\begin{bmatrix}
F_x \\ F_y \end{bmatrix}_{P_0}
 = 
{\begin{bmatrix}
\sum \Delta x_i^2 &  \sum \Delta x_i \Delta y_i \\
\sum \Delta x_i \Delta y_i & \sum \Delta y_i^2
\end{bmatrix}}^{-1} 
\begin{bmatrix}
\sum \Delta x_i \Delta \widetilde{ F}_i \\
\sum \Delta y_i \Delta \widetilde{ F}_i \\
\end{bmatrix}_{P_i \in N \left( P_0 \right) }
\label{ls-formulae-fx-fy-second-order}
\end{equation}
Taking $ \boldsymbol{\Psi} $ - moments of eq. (\ref{kfvs-2d-be}) along with the above least-squares formulae, we get the semi-discrete second-order upwind meshfree scheme based on LSKUM for $2D$ Euler equations,
\begin{equation}
\frac{d \boldsymbol{U}}{d t} + \frac{\partial \boldsymbol{Gx}^+}{\partial x } + \frac{\partial \boldsymbol{Gx}^-}{\partial x } + \frac{\partial \boldsymbol{Gy}^+}{\partial y } + \frac{\partial \boldsymbol{Gy}^-}{\partial y }  = 0 
\label{kfvs-ee}
\end{equation}
Here, $\boldsymbol{Gx}^{\pm}$ and $\boldsymbol{Gy}^{\pm}$ are the kinetic split fluxes \cite{lskum-ghosh-journal} along the $x$ and $y$ directions, respectively. The expressions for the spatial derivatives of $\boldsymbol{Gx}^{\pm}$ are given by 
\begin{equation}
\frac{\partial \boldsymbol{Gx}^{\pm}}{\partial x } 
 = \frac{\sum \Delta y_i^2 \sum \Delta x_i \Delta \widetilde{\boldsymbol{Gx}}_i^{\pm} - \sum \Delta x_i \Delta y_i \sum \Delta y_i \Delta \widetilde{\boldsymbol{Gx}}_i^{\pm} }{\sum \Delta x_i^2 \sum \Delta y_i^2  - \left(\sum \Delta x_i \Delta y_i\right)^2 }  
\label{ls-formulae-second-order-gxpm}
\end{equation}
Here, the perturbations $\Delta \widetilde{\boldsymbol{Gx}}_i^{\pm}$ are defined by
\begin{equation}
\Delta \widetilde{\boldsymbol{Gx}}_i^{\pm}=  \Delta {\boldsymbol{Gx}}^{\pm}_i - \frac{1}{2} \left\lbrace   \Delta x_i \frac{\partial}{\partial x} \Delta \boldsymbol{Gx}_i^{\pm} +  \Delta y_i \frac{\partial}{\partial y} \Delta \boldsymbol{Gx}_i^{\pm} \right\rbrace  
\label{pert-mod-split-fluxes}
\end{equation}
The split flux derivatives in eq. (\ref{ls-formulae-second-order-gxpm}) are approximated using the stencils $\scalemath{0.9}{ N_x^{\pm} \left( P_0 \right)  =   \left\lbrace P_i \in N \left( P_0 \right) \mid \Delta x_i \lessgtr 0 \right\rbrace}$. Similarly, we can write the formulae for the spatial derivative of $\boldsymbol{Gy}^{\pm}$. An advantage of the defect correction procedure is that the least-squares formulae for the second-order approximations to the spatial derivatives of split fluxes are similar to the first-order formulae. For example, to get first order formulae for the spatial derivatives of ${\boldsymbol{Gx}}^{\pm}$, $\Delta \widetilde{\boldsymbol{Gx}}_i^{\pm}$ in eq. (\ref{pert-mod-split-fluxes}) are replaced by $\Delta {\boldsymbol{Gx}}_i^{\pm}$. However, a drawback of the defect correction approach using the Maxwellian distributions is that the numerical solution may not be positive as $\Delta \widetilde{F}_i$ is not the difference between two Maxwellians. Instead, it is the difference between two perturbed Maxwellian distributions $\widetilde{F}_i$ and $\widetilde{F}_0$ \cite{qlskum}.  For preserving the positivity of the solution, instead of the Maxwellians, $\boldsymbol{q}$-variables \cite{smd-nasa-report-entropy-variables} can be used in the defect correction procedure \cite{qlskum}. Note that $\boldsymbol{q}$-variables can represent the fluid flow at the macroscopic level as the transformations $\boldsymbol{U} \longleftrightarrow \boldsymbol{q} $ and $F\longleftrightarrow \boldsymbol{q}$ are unique. The $\boldsymbol{q}$-variables in $2D$ are defined by
\begin{equation}
\boldsymbol{q} = \begin{bmatrix}
\ln \rho + \frac{\ln \beta}{\gamma - 1} - \beta \left(u_1^2 + u_2^2 \right), \; 2 \beta u_1, \; 2 \beta u_2, \; -2\beta 
\end{bmatrix}, \; \beta = \frac{\rho}{2 p}    
\label{q_variables}
\end{equation}
where $\rho$ is the fluid density, $p$ is the pressure and $\gamma$ is the ratio of the specific heats. Using $\boldsymbol{q}$-variables, a second-order meshfree scheme can be obtained by replacing $\Delta \widetilde{\boldsymbol{Gx}}_i^{\pm}$ in eq. (\ref{pert-mod-split-fluxes}) with $\Delta \boldsymbol{Gx}_i^{\pm} \left(\widetilde{\boldsymbol{q}}\right) =  \boldsymbol{Gx}^{\pm} \left(\widetilde{\boldsymbol{q}}_i\right) - \boldsymbol{Gx}^{\pm} \left(\widetilde{\boldsymbol{q}}_0\right) 
$. Here, $\widetilde{\boldsymbol{q}}_i$ and $\widetilde{\boldsymbol{q}}_0$ are the modified $\boldsymbol{q}$-variables, defined by
\begin{equation}
\widetilde{\boldsymbol{q}}_{i}  =  \boldsymbol{q}_{i} - \frac{1}{2} \left( \Delta x_i {\boldsymbol{q}_x}_{i} +  \Delta y_i {\boldsymbol{q}_y}_{i}  \right), \; \; \widetilde{\boldsymbol{q}}_{0}  =  \boldsymbol{q}_{0} - \frac{1}{2} \left( \Delta x_i {\boldsymbol{q}_x}_{0} +  \Delta y_i {\boldsymbol{q}_y}_{0}  \right) 
\label{q-tilde-variables}
\end{equation}
Here $\boldsymbol{q}_x$ and $\boldsymbol{q}_y$ are evaluated to second-order using the least-squares formuale
\begin{equation}
    \begin{bmatrix}
    \boldsymbol{q}_x \\ \boldsymbol{q}_y
    \end{bmatrix} = {\begin{bmatrix}
\sum \Delta x_i^2 &  \sum \Delta x_i \Delta y_i \\
\sum \Delta x_i \Delta y_i & \sum \Delta y_i^2  
\end{bmatrix}}^{-1} 
\begin{bmatrix}
\sum \Delta x_i  \Delta \widetilde{ \boldsymbol{q}}_i  \\
\sum \Delta y_i \Delta \widetilde{\boldsymbol{q}}_i  \\
\end{bmatrix}_{P_i \in N \left( P_0 \right) } 
\label{ls-formulae-q-derivatives}
\end{equation}
The above formulae for $\boldsymbol{q}_x$ and $\boldsymbol{q}_y$ are implicit and need to be solved iteratively. These iterations are known as inner iterations. Finally, the state-update formula for eq. (\ref{kfvs-ee}) can be constructed using a suitable time marching scheme for the transient term with local time stepping \cite{kfvs}. Using the first-order forward difference formula, the state-update for the steady-state flow problems is given by 
\begin{equation}
\boldsymbol{U}^{n+1} = \boldsymbol{U}^{n} - \Delta t  \left\lbrace \frac{\partial \boldsymbol{Gx}^+}{\partial x } + \frac{\partial \boldsymbol{Gx}^-}{\partial x } + \frac{\partial \boldsymbol{Gy}^+}{\partial y } + \frac{\partial \boldsymbol{Gy}^-}{\partial y } \right\rbrace
\label{state-update}
\end{equation}
\section{Regent based Meshfree LSKUM Solver}
\label{sec-regent-lskum-solver}

Algorithm {\ref{algo-serial}} presents a general structure of the serial meshfree LSKUM solver for steady flows \cite{regent-paper-lskum}. The solver consists of a fixed point iterative scheme, denoted by the {\tt for} loop. The subroutine {\tt q\_variables()} computes the $\boldsymbol{q}$-variables defined in eq. (\ref{q_variables}). The routine {\tt q\_derivatives()} evaluates the second-order accurate $\boldsymbol{q}_x$ and $\boldsymbol{q}_y$ using the least-squares formulae in eq. (\ref{ls-formulae-q-derivatives}). {\tt flux\_residual()} computes the sum of kinetic split flux derivatives in (\ref{kfvs-ee}). {\tt timestep()} finds the local time step $\Delta t$ in eq. (\ref{state-update}). Finally, {\tt state\_update()} updates the flow solution using the formula in eq. (\ref{state-update}). The subroutine {\tt residue()} computes the $L_2$ norm of the residue. All the input operations are performed in {\tt preprocessor()}, while the output operations are in {\tt postprocessor()}. The parameter $N$ represents the fixed point iterations required to achieve a desired convergence in the flow solution. 
\begin{algorithm}[t]
    \DontPrintSemicolon
    \SetAlgoLined
    \vspace{1mm}
    \SetKwFunction{FMain}{LSKUM}
    \SetKwProg{Fn}{subroutine}{}{}
        \Fn{\FMain}
    {
                \vspace{1mm}
                call {\tt{ preprocessor() }}\\
                \vspace{1mm}
        \For{$t \leftarrow 1$ \KwTo $t \leq N$}
        {
            \vspace{1mm}
                         call {\tt  q\_variables() } \\
                \vspace{1mm}
                         call    {\tt  q\_derivatives() } \\
                \vspace{1mm}
                         call {\tt  flux\_residual() } \\
                \vspace{1mm}
                     call {\tt{timestep() }}\\
                \vspace{1mm}
                         call  {\tt state\_update() } \\
                \vspace{1mm}
                        call {\tt  residue() } \\
                        \vspace{1mm}
        }
                \vspace{1mm}
                call {\tt{ postprocessor() }}\\
                \vspace{1mm}
    }
    \vspace{1mm}
    \textbf{end subroutine}
    \vspace{1mm}
    \caption{Serial meshfree solver based on LSKUM}
    \label{algo-serial}
\end{algorithm}
%
%
%

\subsection{GPU Accelerated Solver}
Listing \ref{cuda-solver} shows the LSKUM function written in CUDA-C. This code consists of the following sequence of operations: transfer the input data from the host ({\tt{point\_h}}) to the device ({\tt{point\_d}}), perform the fixed point iterations on the device, and transfer the converged flow solution from device to host. Note that all the pre-processing and post-processing operations are performed on the host. In CUDA-C, each function inside the fixed point iteration must be converted into equivalent CUDA kernels, as shown in Listing \ref{cuda-solver}. Furthermore, the developer must handle the memory operations explicitly using pointers to the input point distribution. To launch a CUDA-C kernel, the developer must provide the execution configuration. The execution configuration consists of the number of CUDA threads per block and the total number of blocks. For optimal speedup, the user needs to fine-tune the execution configuration. Note that the CUDA-C code can run only on NVIDIA GPUs but not AMD GPUs. For AMD GPUs, the developer has to rewrite the code using the ROCm framework \cite{AMD_ROCm}. \\ \\
    \begin{lstlisting}[language=CUDA C, caption= LSKUM function written in CUDA-C., label=cuda-solver]
void lskum_cuda(points* point_d, cudaStream_t stream)
{
    preprocessor();
    double residue = 0.0;
    cudaMemcpy(point_d, point_h, sizeof(points), cudaMemcpyHostToDevice, stream);
    for (it = 1; it <= N  ; it++)
    {
        q_variables_cuda<<<blocksPerGrid, threadsPerBlock>>>(*point_d);   
        q_derivatives_cuda<<<blocksPerGrid, threadsPerBlock>>>(*point_d);         
        flux_residual_cuda<<<blocksPerGrid, threadsPerBlock>>>(*point_d);
        timestep_cuda<<<blocksPerGrid, threadsPerBlock>>>(*point_d);
        state_update_cuda<<<blocksPerGrid, threadsPerBlock>>>(*point_d);
        residue = calculate_iteration_residue(*point_d);
    }
    cudaMemcpy(&point_h, point_d, sizeof(double), cudaMemcpyDeviceToHost, stream);
    postprocessor();
}
\end{lstlisting}
 Listing \ref{regent-solver} shows the LSKUM task written in Regent. Note that the Regent code shown in Listing \ref{regent-solver} can run on any heterogeneous platform consisting of GPUs (NVIDIA or AMD) and CPUs. If the user requests a GPU to be used, the operations performed are largely the same as the CUDA-C solver: the input data is transferred from host to device, perform the fixed point iterations on the device, and the converged flow solution is transferred back to the host. However, the user does not need to write the memory transfer operations, unlike in CUDA-C. An advantage of Regent is that instead of the user, the compiler automatically generates a kernel that can run on any GPU. Furthermore, the Regent compiler implicitly sets the execution configuration for kernel launches. \\ \\
    \begin{lstlisting}[language=Regent, caption=LSKUM task written in Regent., label=regent-solver]
task lskum(points : region(ispace(int1d), point)) 
where reads writes (points)
do
    preprocessor()
    var residue : double = 0.0
    var points_allnbhs = local_points | ghost_points
    for t = 1, N do
        for i = 1, local_points.colors do
            q_variables(local_points[i])
        end
        for i = 1, local_points.colors do
            q_derivatives(local_points[i], points_allnbhs[i])
        for i = 1, local_points.colors do
            flux_residual(local_points[i], points_allnbhs[i])
        end
        for i = 1, local_points.colors do
            local_time_step(local_points[i], points_allnbhs[i])
        end
        for i = 1, local_points.colors do
            residue += state_update(local_points[i])
        end
    end
    postprocessor()
end
\end{lstlisting}
%
%
%
We demonstrate the changes a developer has to make to write a CUDA-C kernel through an example.
Listing \ref{lst:example-serial-C} shows the serial code in C to compute the {\tt q\_variables} at all points in the computational domain, and Listing \ref{lst:converted-CUDA-Code} shows the corresponding CUDA-C kernel. 
In CUDA-C, the developer must convert the for-loop in the serial code into an equivalent if-statement and ensure that the {\tt thread\_index} is calculated correctly. In this example, the {\tt thread\_index} must lie between $0$ and the total points in the input point distribution. Note that the total number of CUDA threads must equal to or exceed the maximum points. 
\begin{lstlisting}[language=C, caption=Serial code in C to compute $\boldsymbol{q}$-variables at all points in the computational domain. ,  label=lst:example-serial-C]
void q_variables()
{
    double rho, u1, u2, pr, beta;
    for (int k = 0; k < max_points; k++)
    {
        rho = point.prim[0][k];
        u1 = point.prim[1][k];
        u2 = point.prim[2][k];
        pr = point.prim[3][k];
        beta = 0.5*r/pr;
        point.q[0][k] = log(rho)+(log(beta)*2.5)-beta*(u1 * u1 + u2 * u2);
        point.q[1][k] = 2.0* beta * u1;
        point.q[2][k] = 2.0* beta * u2;
        point.q[3][k] = -2.0 * beta;
    }
}      
\end{lstlisting}    
\begin{lstlisting}[language=CUDA C, caption= CUDA-C code for computing $\boldsymbol{q}$-variables at all points in the computational domain. , label=lst:converted-CUDA-Code]
__global__ void q_variables_cuda(points &point)
{
    double rho, u1, u2, pr, beta;
    int bx = blockIdx.x;
    int tx = threadIdx.x;
    int thread_index = blockIdx.y * gridDim.x + blockIdx.x * blockDim.x + threadIdx.x;
    if (thread_index < 0 || thread_index >= max_points){
        return;
    }
    rho = point.prim[0][thread_index];
    u1 = point.prim[1][thread_index];
    u2 = point.prim[2][thread_index];
    pr = point.prim[3][thread_index];
    beta = 0.5 * rho / pr;
    point.q[0][thread_index] = log(rho) + (log(beta) * 2.5) - beta * (u1 * u1 + u2 * u2);
    point.q[1][thread_index] = 2.0 * beta * u1;
    point.q[2][thread_index] = 2.0 * beta * u2;
    point.q[3][thread_index] = - 2.0 * beta;
} 
\end{lstlisting}
The portion of the Regent code that computes {\tt q\_variables} is shown in Listing \ref{lst:example-serial-regent}. In order to execute this task on a GPU, the user needs to add {\tt \_\_demand(\_\_cuda)} annotation above the task definition as shown in Listing \ref{lst:example-cuda-regent}.  Note that the other tasks in Listing \ref{regent-solver} can be executed on the GPU in a similar fashion. However, if there is no GPU in the system configuration, this task will run on a CPU. \\ 
\begin{lstlisting}[language=Regent, caption=Regent task for computing $\boldsymbol{q}$-variables at all points in the computational domain. , label=lst:example-serial-regent]
task q_variables(points : region(ispace(int1d), point)) where
writes (points.{q}),
reads (points.{prim})
do
    for point in points do
        var rho : double = point.prim[0]
        var u1 : double = point.prim[1]
        var u2 : double = point.prim[2]
        var pr : double = point.prim[3]
        var beta : double = 0.5*rho/pr
        point.q[0] = log(rho) + log(beta)*2.5 - beta*(u1*u1 + u2*u2)
        point.q[1] = 2.0*beta*u1 
        point.q[2] = 2.0*beta*u2
        point.q[3] = - 2.0*beta 
    end
end
\end{lstlisting}
\begin{lstlisting}[language=Regent, caption=Regent task on a GPU for computing $\boldsymbol{q}$-variables at all points in the computational domain., label=lst:example-cuda-regent]
__demand(__cuda)
task q_variables(points : region(ispace(int1d), point)) where
writes (points.{q}),
reads (points.{prim})
do
    for point in points do
        var rho : double = point.prim[0]
        var u1 : double = point.prim[1]
        var u2 : double = point.prim[2]
        var pr : double = point.prim[3]
        var beta : double = 0.5*rho/pr
        point.q[0] = log(rho) + log(beta)*2.5 - beta*(u1*u1 + u2*u2)
        point.q[1] = 2.0*beta*u1 
        point.q[2] = 2.0*beta*u2
        point.q[3] = - 2.0*beta 
    end
end
\end{lstlisting}

%
%
\subsection{CPU Parallel Solver}
 In general, parallelising a meshfree LSKUM solver using MPI on CPUs involves two steps. In the first step, the given domain is decomposed into smaller domains, known as partitions. Later, suitable subroutines are written by the user for data communication and synchronisation between the partitions. In the present CPU parallel LSKUM code based on Fortran, the domain, which is an unstructured point distribution, is partitioned using METIS \cite{metis}. \\  \\
 The points in a decomposed partition are classified into two groups, namely, the local points and ghost points. Local points are those points that are contained within the partition. For the local points near the boundary of the partition, some of the neighbours in the connectivity set lie in other partitions. The collection of these neighbouring points that lie in other partitions are known as the ghost points for the current partition. \\ \\ 
%
%
%
In the CPU parallel code based on Fortran, the computation of {\tt q\_variables} and {\tt state\_update} in Algorithm \ref{algo-serial} at the local points in each partition does not require any information from other partitions. However, the computation of {\tt q\_derivatives} and {\tt flux\_residual} at a point in a given partition requires the updated values of $\boldsymbol{q}$-variables and $\boldsymbol{q}$-derivatives at its connectivity. In order to get the updated values for the ghost points and synchronise the data, the code uses the communication calls MPI SendRecv and MPI Barrier. Furthermore, the code uses MPI Allreduce to compute the residue and aerodynamic force coefficients, such as lift and drag coefficients. Note that all the communication calls are handled using PETSc \cite{petsc} libraries. \\ \\
%
 %
%
%
In Regent, the domain is stored in a data structure known as a region. In our code, the region consists of the distribution of points in the computational domain. METIS is then used to assign a colour to each point in the region. Points with the same colour are grouped into subregions using Regent's inbuilt partitioning system. The points in a subregion are known as local points, shown in Listing \ref{regent-solver}. Note that the number of subregions is equivalent to the number of CPU cores used. Using the inbuilt partitioning system, the user has to create a set of neighbouring points for all the local points in the subregion \cite{regent-paper-lskum}. This set contains both the local and ghost points for that subregion. In Listing \ref{regent-solver}, we refer to this set as {\tt points\_allnbhs}. \\ \\
Unlike Fortran, Regent does not use MPI for data communication and synchronisation. Instead, it is handled implicitly by Realm, a low-level runtime system \cite{Realm14}. This system handles all the communications using the GASnet networking layer \cite{GASNET07} while, the Legion runtime system takes care of the synchronisation \cite{Legion12}. Finally, the residue and aerodynamic force coefficients are computed using a reduction operator. Regent replaces this operator with the optimal runtime call. 
\section{Verification of the Regent Solver } 
\label{sec-code-verification}
\begin{figure}[htbp]
\centering
    \subfloat[\centering Mach contours. ]{{\includegraphics[width=0.48\textwidth,trim={5mm 5mm 5mm 5mm},angle=0,clip]{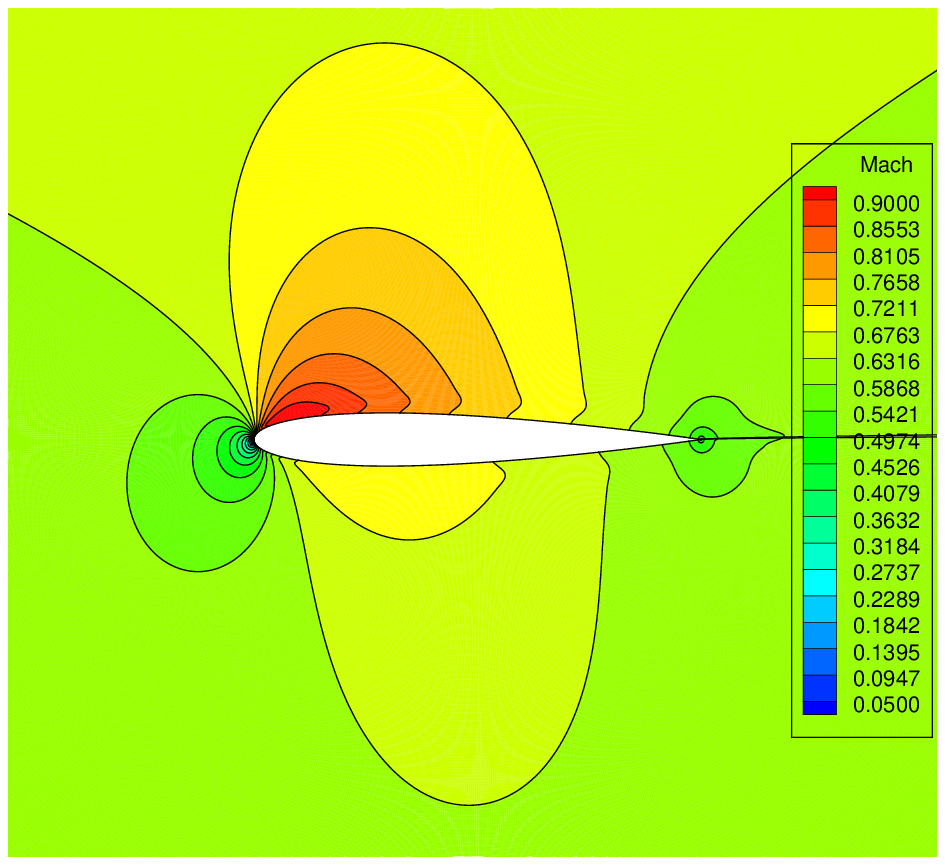} }}
        \subfloat[\centering Surface pressure distribution.]{{\includegraphics[width=0.48\textwidth,trim={5mm 5mm 5mm 5mm},angle=0,clip]{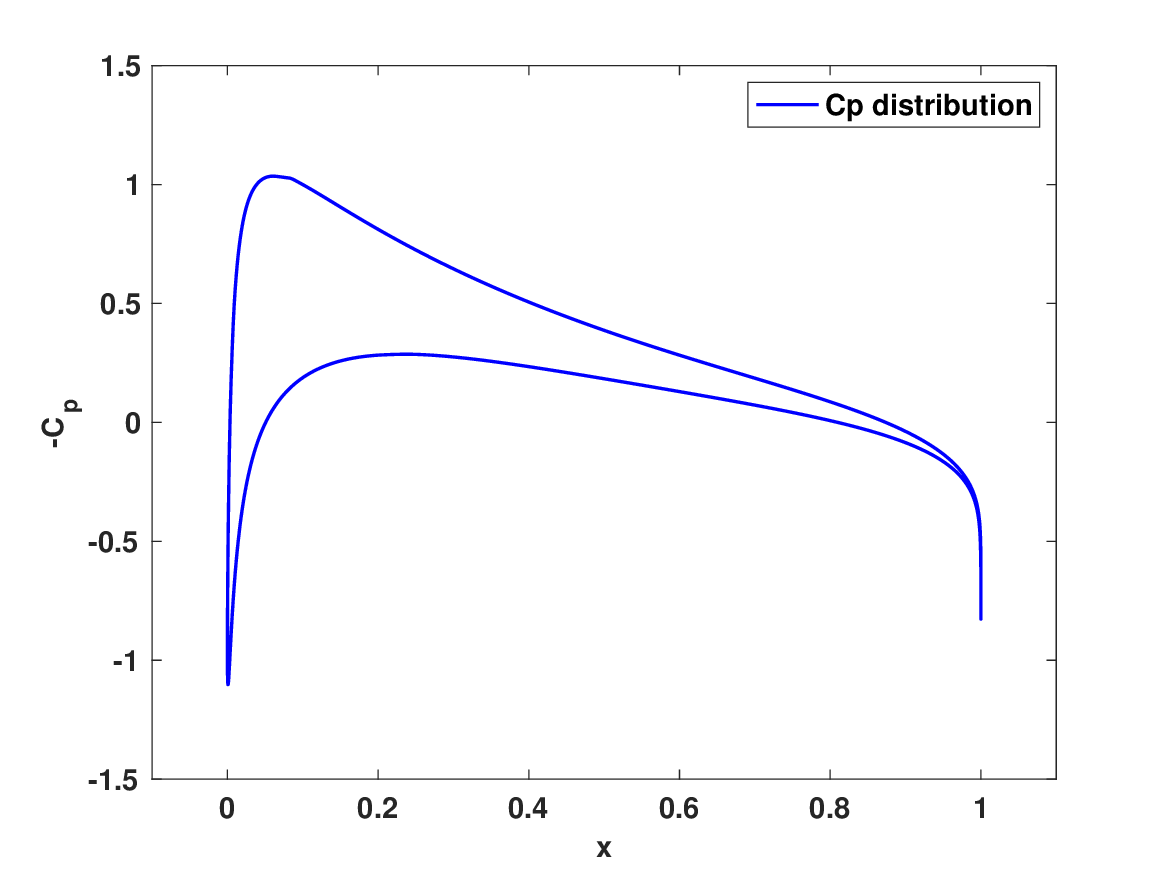} }}
\caption{Subsonic flow around the NACA 0012 airfoil at $M_{\infty} = 0.63 $ and $AoA = 2^o$.  }
\label{subsonic-case}
 \end{figure}

\begin{figure}[htbp]
\centering
    \subfloat[\centering Pressure contours. ]{{\includegraphics[width=0.48\textwidth,trim={5mm 5mm 5mm 5mm},angle=0,clip]{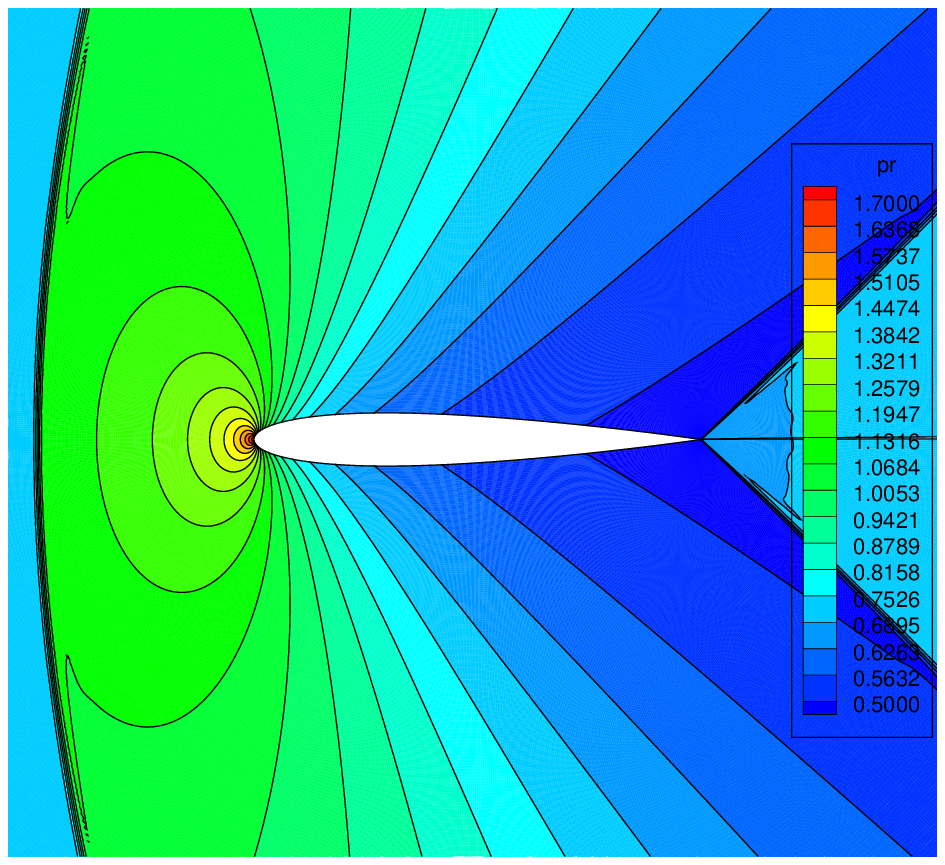} }}
        \subfloat[\centering Surface pressure distribution.]{{\includegraphics[width=0.48\textwidth,trim={5mm 5mm 5mm 5mm},angle=0,clip]{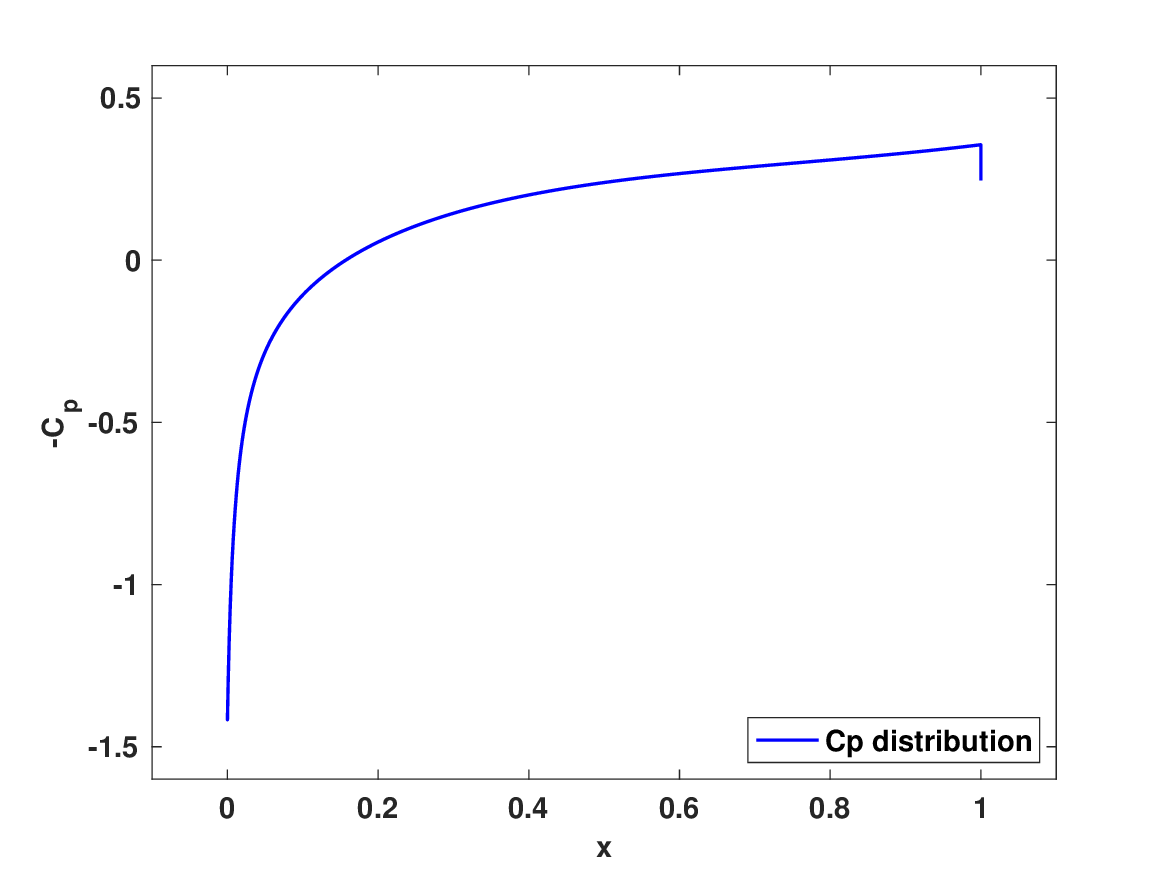} }}
\caption{Supersonic flow around the NACA 0012 airfoil at $M_{\infty} = 1.2 $ and $AoA = 0^o$.  }
\label{supersonic-case}
 \end{figure}

To verify the Regent based meshfree LSKUM solver, numerical simulations are performed on the NACA 0012 airfoil at subsonic and supersonic flows. For the subsonic case, the freestream conditions are given by the Mach number, $M_{\infty} = 0.63$ and the angle of attack, $AoA = 2^o$. For the supersonic case, the flow conditions are $M_{\infty} = 1.2$ and $AoA = 0^o$. The computational domain consists of $614,400$ points. The airfoil is discretised with $1280$ points. Figures \ref{subsonic-case} and \ref{supersonic-case} show the flow contours and the surface pressure distribution on the airfoil. These plots show that the Regent solver accurately captures the desired flow features, such as the suction peak in the subsonic case and the bow and fish tail shocks in the supersonic case. Furthermore, the computed flow solution matches up to machine precision with the solutions obtained from the corresponding serial code in Fortran and CUDA-C GPU code \cite{hipc-paper}.
\section{Performance Analysis of the Regent Solver on a GPU } 
\label{sec-per-analysis-gpu}
\begin{figure}[htbp]
{\centering
\includegraphics[scale=0.50,trim={0 0 10mm 5mm},angle=0,clip]{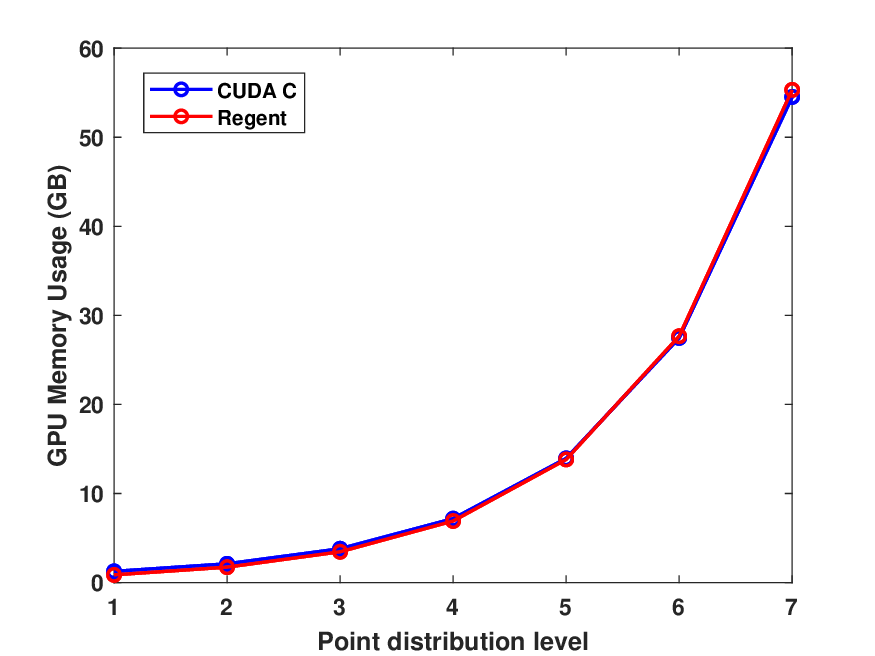}
\caption{ GPU memory used by the Regent and CUDA-C codes. }
     \label{memory-naive-codes}
}
\end{figure}

This section presents the numerical results to assess the performance of the Regent based meshfree LSKUM solver on a single GPU card. We compare the performance of the code with an optimised CUDA-C implementation of the same solver. In \cite{DBLP:journals/corr/abs-2108-07031}, it was shown that the CUDA-C based LSKUM solver exhibited superior performance over the equivalent GPU solvers written in Fortran, Python, and Julia. In the present work, the CUDA-C code is compiled with \texttt{nvcc 22.5} using the flags: \texttt{-O3} and \texttt{-mcmodel=large} \cite{cuda}. Table \ref{workstation-config} shows the hardware configuration used for the simulations. \\ \\ 
 For the benchmarks, numerical simulations are performed on seven levels of point distributions around the NACA 0012 airfoil at Mach number, $M = 0.63$ and angle-of-attack, $AoA = 2^{\circ}$. The coarsest distribution consists of one million points, while the finest distribution has $64$ million points. The GPU memory used by Regent and CUDA-C codes for these point distributions is shown in Figure \ref{memory-naive-codes}. We observe that the memory usage of both codes is nearly the same. 
\begin{table}[htbp]
\centering
\begin{tabular}{lcc}
\toprule
& CPU   & GPU   \\
\midrule
Model & AMD EPYC\textsuperscript{TM} $ 7532 $   & Nvidia Ampere $A100$ SXM4  \\ [0.3em]
Cores & $64$ $\left( 2 \times 32 \right)$ & $5120$ \\ [0.3em]
Core Frequency & $2.40$ GHz & $1.23$ GHz \\ [0.3em]
Global Memory & $1$ TiB & $80$ GiB \\ [0.3em]
$L2$ Cache & $16$ MiB & $40$ MiB \\ [0.1em]
%
\bottomrule
\end{tabular}
\caption{Hardware configuration of the node with A100 GPU card. }
\label{workstation-config}
\end{table}
\subsection{RDP Comparison of Regent and CUDA-C codes}
\begin{figure}[htbp]
{\centering
\includegraphics[scale=0.50,trim={0 0 10mm 5mm},angle=0,clip]{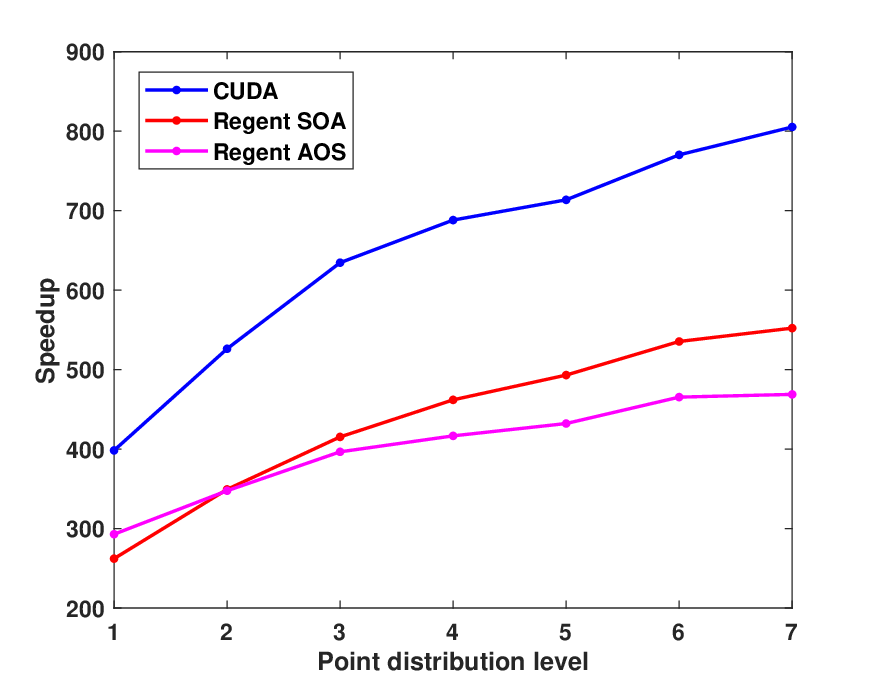}
\caption{ Speedup achieved by the Regent and CUDA-C GPU codes. }
     \label{speedup-naive-codes}
}
\end{figure}

To measure the performance of the codes, we adopt a non-dimensional cost metric called the Rate of Data Processing (RDP). The RDP of a meshfree code can be defined as the total wall clock time in seconds per iteration per point. Note that the lower the RDP, the better \cite{DBLP:journals/corr/abs-2108-07031}. In the present work, the RDP values are measured by specifying the number of pseudo-time iterations in eq. (\ref{state-update}) to $10,000$ and the inner iterations required for second-order accurate approximations of $\boldsymbol{q}_x$ and $\boldsymbol{q}_y$ in eq. (\ref{ls-formulae-q-derivatives}) to $3$. \\ \\ 
Table \ref{rdp-single-node-naive} compares the RDP values for the Regent and CUDA-C meshfree codes. Here, for Regent, we have implemented the solver using two memory storage schemes: Array of Structures (AOS) and Structure of Arrays (SOA). The difference between the AOS and SOA implementations lies in how the memory patterns are accessed. AOS is the conventional memory layout that is supported by most programming languages. In AOS, the fields of a data structure are stored in an interleaved pattern. On the other hand, in SOA, the fields are stored in separate parallel arrays. SOA is preferred for Single Instruction Multiple Data (SIMD) instructions because more data can be packed efficiently into datapaths for processing. \\ \\ 
The tabulated values show that the CUDA-C exhibited superior performance on all levels of point distributions. We also observe that the RDP values of the Regent code with AOS are lower than SOA implementation on coarse distributions. However, on finer point distributions, Regent with SOA yielded lower RDP values. The difference in the performance of the Regent codes can be attributed to the memory access patterns. When consecutive threads access sequential memory locations, it results in multiple memory accesses, which can be combined into a single transaction. This is known as coalesced memory access. The memory access scheme becomes serial if the memory access pattern is non-sequential. This is known as an uncoalesced memory access, leading to a performance loss. In the current implementation, the SOA scheme leads to more coalesced memory accesses than AOS. This is because data related to each field is stored in contiguous memory locations in the SOA scheme. This allows consecutive threads to access sequential memory locations, leading to more coalesced memory accesses. As a result, the Regent-SOA solver has lower RDP values than the Regent-AOS solver. \\ \\ 
To assess the performance of the Regent codes with the CUDA-C code, we define another metric called relative performance. The relative performance of a Regent code is defined as the ratio of the RDP of the Regent code to the CUDA-C code. Table \ref{rdp-single-node-naive} shows that on the finest point distribution, the Regent-AOS code is slower by $1.714$ than the CUDA-C code. On the other hand, the Regent-SOA code is slower by a factor of $1.457$, demonstrating its superior performance over AOS implementation. Figure \ref{speedup-naive-codes} shows the speedup achieved by the GPU codes. Here, the speedup of a GPU code is defined as the ratio of the RDP values of the serial Fortran code to the GPU code. It can be observed that the CUDA-C code achieved a speedup of around $800$. On the other hand, the speedup achieved by Regent-SOA and AOS codes is around $550$ and $450$, respectively. In the following sections, we analyse metrics related to the Regent SOA code, referred to as the baseline Regent GPU code henceforth. 
\begin{table}[t]
\centering
\scalebox{0.90}{
\begin{tabular}{ccccccc}
\toprule
Level & Point & Regent - AOS & Regent - SOA & CUDA-C & \multicolumn{2}{c}{Relative Performance}\\ [0.2em]
\midrule
 \multicolumn{5}{c}{RDP $\times$ $10^{-8}$ (Lower is better)} & \multicolumn{1}{c} {Regent - AOS} & \multicolumn{1}{c} {Regent - SOA} \\ [0.2em]
\cline{1-4} \cline{5-7} \\ [-0.5em]
$1$ & $1$M & $1.151$ & $1.280$ & $0.846$ & $1.360$ & $1.513$\\ [0.4em]
$2$ & $2$M & $0.996$ & $0.984$ & $0.658$ & $1.513$ & $1.495$\\ [0.4em]
$3$ & $4$M & $0.885$ & $0.837$ & $0.553$ & $1.600$ & $1.513$\\ [0.4em]
$4$ & $8$M & $0.849$ & $0.772$ & $0.514$ & $1.651$ & $1.502$\\ [0.4em]
$5$ & $16$M & $0.822$ & $0.720$ & $0.498$ & $1.651$ & $1.445$\\ [0.4em]
$6$ & $32$M & $0.805$ & $0.704$ & $0.487$ & $1.653$ & $1.445$\\ [0.4em]
$7$ & $64$M & $0.799$ & $0.679$ & $0.466$ & $1.714$ & $1.457$\\
\bottomrule
\end{tabular}
}
\caption{A comparison of the RDP values of the Regent and CUDA-C GPU codes.}
\label{rdp-single-node-naive}
\end{table}
\subsection{RDP analysis of kernels}
To understand the performance of the Regent GPU code, we investigate the kernels employed in the solver. Towards this objective, an RDP analysis of the individual kernels is performed. The RDP of a kernel is defined as the runtime of the kernel in seconds per iteration per point. To obtain the kernel runtimes, {\tt NVIDIA NSight Compute} \cite{nvidia-documentation} for CUDA-C and {\tt Legion\_Prof} \cite{legion-documentation} for Regent are used. Note that the implementation of kernels in Regent code is structurally similar to CUDA-C, except for the {\tt flux\_residual} kernel. In CUDA-C, the {\tt flux\_residual} kernel is split into four smaller kernels that compute the spatial flux derivatives of the split fluxes $\boldsymbol{Gx}^+ $, $\boldsymbol{Gx}^- $, $\boldsymbol{Gy}^+ $ and $\boldsymbol{Gy}^- $, in eq. (\ref{state-update}).  The RDP of the {\tt flux\_residual} kernel is then defined as the sum of the RDPs of the smaller kernels. \\ \\ 
Table \ref{naive-gpu-codes-run-time-kernels} shows the RDP of the kernels on coarse ($1$M), medium ($8$M), and finest ($64$M) point distributions. The tabulated values show that the RDP values of the {\tt flux\_residual} and {\tt q\_derivatives} kernels are high and, therefore, contribute significantly to the overall RDP values of the solvers. For the Regent code, the RDP values of the above kernels are higher than CUDA C, resulting in the poor performance of Regent. However, for other kernels with lower RDP values, Regent exhibited slightly superior performance. Note that the RDP of the {\tt q\_derivatives} kernel depends on the number of inner iterations. The higher the inner iterations, the more the RDP of the kernel. In the present work, the RDP values are calculated using three inner iterations. 
\begin{table}[htbp]
\centering
\begin{center}
\scalebox{0.80}{
\begin{tabular}{lcccccc}
\toprule
Points & Code    & q\_variables   & q\_derivatives & flux\_residual & state\_update & local\_time\_step \\[0.2em]
\midrule
\multicolumn{7}{c}{\centering{RDP $\times$ $10^{-10}$ (Lower is better)}}  \\ [0.2em]
\midrule
\multirow{2}{*}{$1$M} & {Regent} & $0.319$ & $40.600$ & $ 67.488$ & $ 0.749 $ & $ 4.314 $\\ [0.4em]
 & {CUDA-C} & $ 0.409 $ & $ 35.686$ & $ 38.063$ & $0.739 $ & $ 4.354 $\\ [0.2em]        
        \cmidrule(lr){1-7}
\multirow{2}{*}{$16$M} & {Regent} & $0.378$ & $22.421 $ & $45.804 $ & $0.756$ & $1.474$\\ [0.4em]
 & {CUDA-C} & $ 0.374$ & $ 17.970$ & $28.815$ & $0.774$ & $1.809$\\ [0.2em]        \cmidrule(lr){1-7}
\multirow{2}{*}{$64$M} & {Regent} & $0.378$ & $21.849$ & $ 47.863$ & $ 0.764$  & $1.383$\\ [0.4em]
& {CUDA-C} & $0.388$ & $ 16.868$ & $ 27.462$ & $0.778$ & $1.729$\\ [0.2em]           \bottomrule
\end{tabular}
}
\end{center}
\caption{{RDP analysis of the kernels in Regent and CUDA-C. }}
\label{naive-gpu-codes-run-time-kernels}
\end{table} 
\subsection{Performance Metrics for the Kernels {\tt q\_derivatives} and {\tt flux\_residual}}
\begin{figure}[H]
{\centering
\begin{minipage}[c]{0.48\textwidth}
\includegraphics[width=\linewidth]{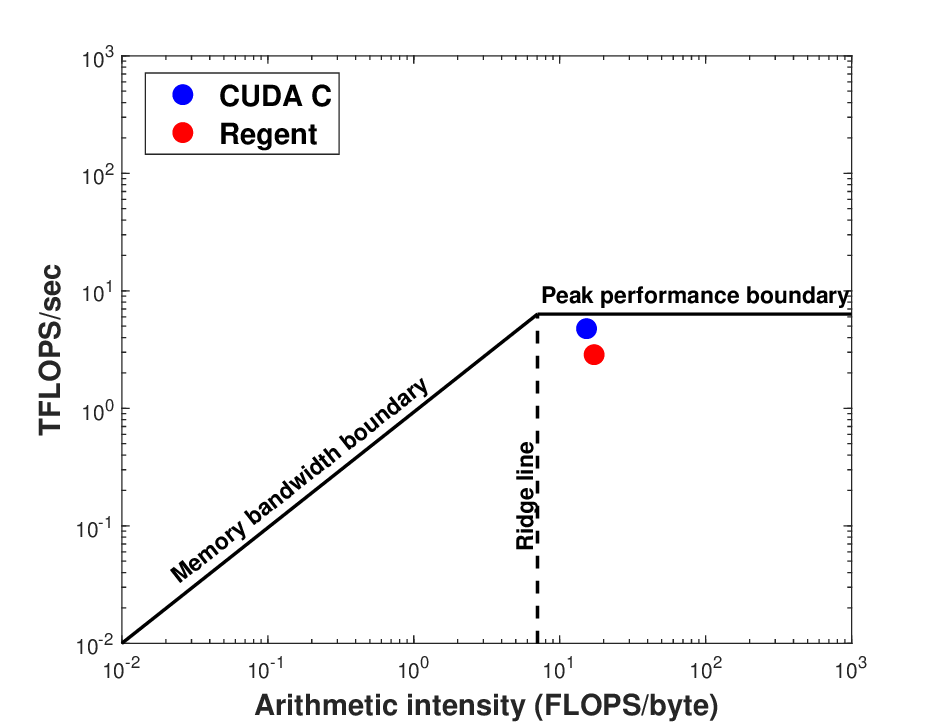}
\caption*{(a) Before splitting the kernel in Regent }
\end{minipage}
\begin{minipage}[c]{0.48\textwidth}
\includegraphics[width=\linewidth]{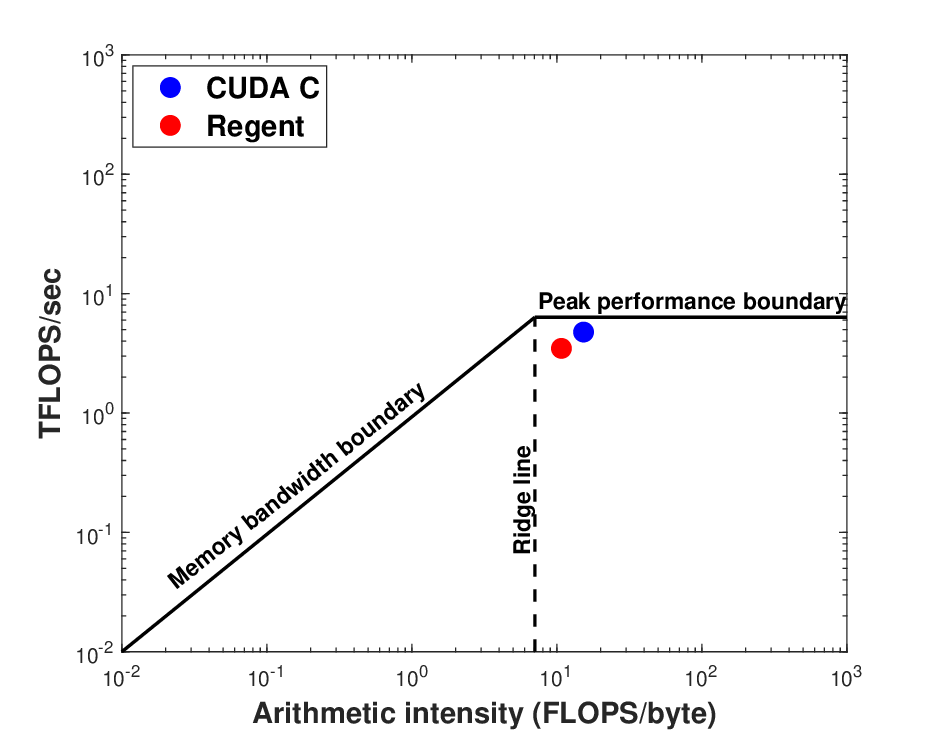}
\caption*{(b) After splitting the kernel in Regent }
\end{minipage}
}
\caption{ \centering Roofline analysis of the Regent and CUDA-C {\tt flux\_residual} kernels.}
\label{roofline-analysis-optimised-codes}
\end{figure}

To understand the reason behind the difference in the RDP values of the Regent and CUDA-C kernels for {\tt q\_derivatives} and {\tt flux\_residual}, we analyse various kernel performance metrics. Table \ref{naive-performance-metrics} shows a comparison of the utilisation of memory, streaming multiprocessors (SM), achieved occupancy and the registers per thread on the coarsest (1M) and fine (32M) levels of point distributions. These statistics are obtained using {\tt NVIDIA NSight Compute} reports. Due to memory constraints, profiling is not performed on the finest point distribution with 64 million points. Since the {\tt flux\_residual} kernel in CUDA-C is split into smaller kernels, we present a range of values for the performance metrics based on split kernels. \\  
\begin{table}[H]
\centering
\begin{center}
\begin{tabular}{cccccc}
\toprule
Points & Code    & Memory & SM   & Achieved  & Registers \\
&  & utilisation &  utilisation  & occupancy & per thread\\[0.2em]
\cmidrule(lr){3-5}
& & \multicolumn{3}{c}{shown in percentage} & \\
\midrule
\multicolumn{6}{c}{{\tt flux\_residual}}\\
\midrule
\multirow{2}{*}{$1$M}& Regent & $41.78  $ & $ 33.38 $  & $11.70 $ & $213$\\ [0.4em]        
 & CUDA-C & $50.55-52.24  $ & $ 63.59-66.67 $  & $16.79-16.87$ & $142$\\ [0.2em]        
       \midrule        
\multirow{2}{*}{$32$M} & Regent & $20.14 $ & $46.11$ &  $ 12.50 $ & $212$\\ [0.4em]
& CUDA-C & $20.39-20.42 $ & $ 73.98-76.81 $ &  $18.10-18.16$ & $144$\\ [0.2em]        
\midrule
\multicolumn{6}{c}{{\tt q\_derivatives}}\\
\midrule
\multirow{2}{*}{$1$M} & Regent & $ 89.29 $ & $ 26.34 $  & $29.51 $ & $93$\\ [0.4em]        
 & CUDA-C & $ 87.47$ & $ 27.64 $  & $23.59$ & $102$\\ [0.2em]        
\midrule        
\multirow{2}{*}{$32$M} & Regent & $ 53.15$ & $53.41$ &  $ 30.28 $ & $93$\\ [0.3em]
 & CUDA-C & $ 56.16$ & $ 69.91 $ &  $29.78$ & $96$\\ [0.1em]        
\bottomrule
\end{tabular}
\end{center}
\caption{{A comparison of performance metrics on coarse and fine point distributions. }}
\label{naive-performance-metrics}
\end{table}
We first analyse the memory utilisation of the GPU codes. This metric shows the peak usage of device memory pipelines. 
The closer the memory utilisation is to the theoretical limit, the stronger the possibility that it can be a bottleneck in the performance of the code. However, low memory utilisation does not necessarily imply better performance \cite{nvidia-kernel-profiling-guide}. 
The theoretical limit of memory utilisation can be reached if - 
\begin{enumerate}[label=(\alph*)]
    \item The memory hardware units are fully utilised.
    \item The communication bandwidth between these units is saturated.
    \item The maximum throughput of issuing memory instructions is achieved.
\end{enumerate}
The tabulated values show that the Regent and CUDA-C have high memory utilisation on the coarse distribution. This resulted in higher RDP values for both kernels. However, continuous refinement in the point distribution resulted in a more balanced usage of memory resources, reducing the RDP values of these kernels. \\ \\ 
From Table \ref{naive-performance-metrics}, we also observe that the CUDA-C code has a higher SM utilisation. High utilisation is an indicator of efficient usage of CUDA streaming multiprocessors (SM), while lower values indicate that GPU resources are under-utilised. For the Regent code, the relatively poor utilisation of SM resources resulted in higher RDP values for both kernels. To understand the poor utilisation of the SM resources in Regent, we investigate the achieved occupancy of the kernels. The achieved occupancy is the ratio of the number of active warps to the maximum number of theoretical warps per SM. A code with higher occupancy can allow the SM to execute more active warps, which may increase SM utilisation. Similarly, low occupancy may result in lower SM utilisation. For the Regent code, the {\tt flux\_residual} kernel has low occupancy, which resulted in poor utilisation of SM resources compared to CUDA-C. On the other hand, the achieved occupancy of the {\tt q\_derivatives} kernel is higher, but its SM utilisation is still lower. \\ \\ 
To understand why the {\tt flux\_residual} kernel in Regent has lower occupancy, we analyse its register usage. The higher the register usage, the fewer the active warps. From Table \ref{naive-performance-metrics}, we observe that the {\tt flux\_residual} kernel in Regent has higher register usage, which resulted in lower occupancy and hence, lower SM utilisation. To reduce the register pressure, the {\tt flux\_residual} kernel is split into four smaller kernels. Similar to CUDA-C, these kernels are employed to compute the spatial derivatives of the split fluxes $\boldsymbol{Gx}^+ $, $\boldsymbol{Gx}^- $, $\boldsymbol{Gy}^+ $ and $\boldsymbol{Gy}^- $. Note that the sum of these split flux derivatives yields the {\tt flux\_residual}. \\ \\  
To visualise the performance improvement of the {\tt flux\_residual} kernel in {\tt Regent}, we present its roofline chart before and after kernel splitting. A roofline chart \cite{roofline} is a logarithmic plot that shows a kernel's arithmetic intensity with its maximum achievable performance. The arithmetic intensity is defined as the number of floating-point operations per byte of data movement. The achieved performance is measured in trillions of floating-point operations per second. 
Figure \ref{roofline-analysis-optimised-codes} shows the roofline charts before and after splitting the {\tt flux\_residual} kernel. For this comparison, we define the achieved performance and arithmetic intensity of the {\tt flux\_residual} kernel after splitting as the averaged values of the metrics for the split kernels. Note that for CUDA-C, the metrics shown are the averaged values for the split kernels. After splitting, we observe that the kernel performance in Regent is less compute-bound and has moved closer to the peak performance boundary. This behaviour can be attributed to the more efficient scheduling of the warps by the warp schedulers. To begin the kernel execution on the SM, the unsplit {\tt flux\_residual} kernel requires a significant amount of resources, such as memory, arithmetic units, and registers. By splitting the {\tt flux\_residual} kernel into smaller kernels, the required resources are observed to be considerably lower. This allowed the scheduler to execute more kernels simultaneously on the SM. \\  
Table \ref{naive-performance-metrics-2} shows the performance metrics before and after splitting the {\tt flux\_residual} kernel. Similar to CUDA-C, for the Regent code with split kernels, we present a range of values for the metrics based on split kernels. After splitting the kernel, we observe that the number of registers per thread has reduced. This improved the achieved occupancy of Regent and is comparable to CUDA-C. Although splitting the kernels improved the SM utilisation, it is still much lower than CUDA-C. Due to splitting, the data is transferred over a shorter period, leading to increased activity in the memory pipelines. This resulted in higher utilisation of the memory units. \\ 
\begin{table}[H]
\centering
\scalebox{0.90}{
\begin{tabular}{cccccc}
\toprule
Points & Code    & Memory & SM   & Achieved  & Registers \\
&  & utilisation &  utilisation  & occupancy & per thread\\[0.2em]
\cmidrule(lr){3-5}
& & \multicolumn{3}{c}{shown in percentage} & \\
\midrule
\multicolumn{6}{c}{{\tt flux\_residual}}\\
\midrule
\multirow{3}{*}{$1$M} &  \textcolor{red}{ {\tt Regent - Baseline}} & \textcolor{red}{$41.78$} & \textcolor{red}{$33.38$} & \textcolor{red}{$11.70$} & \textcolor{red}{$213$ } \\ [0.4em]
&   {\tt Regent with split kernels} & $ 52.82-55.92$ & $ 35.51-37.25 $  & $17.07-17.09 $ & $164$\\ [0.4em]        
 & CUDA-C & $ 50.55-52.24$ & $ 63.59-66.67 $  & $16.79-16.87$ & $142$\\ [0.2em]        
       \midrule        
\multirow{3}{*}{$32$M} & \textcolor{red}{ {\tt Regent - Baseline}} & \textcolor{red}{ $20.14$} & \textcolor{red}{ $46.11$} & \textcolor{red}{ $12.50$} & \textcolor{red}{ $212$ }\\ [0.4em] 
   & {\tt Regent with split kernels} & $ 34.86-36.81$ & $52.62-56.06$ &  $ 17.51-17.71 $ & $164$\\ [0.4em]
 & CUDA-C & $ 20.42-20.52$ & $ 73.98-76.81 $ &  $18.10-18.16$ & $144$\\ [0.2em]               
\bottomrule
\end{tabular}
        }
\caption{{A comparison of performance metrics based on optimised Regent code on coarse and fine point distributions. }}
\label{naive-performance-metrics-2}
\end{table}
To understand the low SM utilisation of Regent over CUDA-C, we analyse the scheduler statistics for the {\tt flux\_residual} and {\tt q\_derivatives} kernels. A scheduler maintains warps for which it can issue instructions. Scheduler statistics consist of the following metrics -  {\it active}, {\it eligible} and {\it issued} warps \cite{DBLP:journals/corr/abs-2108-07031}. {\it Active} warps are warps for which resources such as registers and shared memory are allocated. {\it Eligible} warps are {\it active} warps that have not been stalled and are ready to issue an instruction. A subset of {\it eligible} warps, known as {\it issued} warps, denotes the warps selected by the scheduler for execution. The number of {\it active} warps is the sum of the {\it eligible} and {\it stalled} warps \cite{DBLP:journals/corr/abs-2108-07031}.
\begin{table}[htbp]
        \centering
        \begin{center}
        \begin{tabular}{ccccccc}
            \toprule
            \multicolumn{1}{c}{Points} & \multicolumn{1}{c} {Code} & Active & Eligible & Stalled & Issued &Eligible warps\\
\cmidrule(lr){3-6}
  & & \multicolumn{4}{c}{warps per scheduler  } & \multicolumn{1}{c}{in percentage}\\
\midrule
\multicolumn{7}{c}{\tt{flux\_residual  }} \\
\midrule
\multirow{2}{*}{$1$M} & Regent & $2.74 $ & $ 0.24-0.25 $ & $2.49-2.50$ & $0.20-0.21 $ & $19.75-20.89 $\\ [0.4em]  & CUDA-C & $2.69-2.70 $ & $ 0.39-0.42 $ & $2.28-2.30$ & $0.29-0.30$ & $29.00-30.37$ \\ [0.2em]        
       \midrule        
\multirow{2}{*}{$32$M}& Regent & $2.80-2.83$ & $0.38-0.42 $ & $2.41-2.42$ & $0.29-0.31$ & $29.28-31.19 $ \\ [0.4em]
 & CUDA-C & $2.90 $ & $ 0.53-0.49 $ & $2.37-2.41$ & $0.34-0.36$ & $34.21-35.53 $ \\ [0.2em]               
        \midrule
\multicolumn{7}{c}{\tt{q\_derivatives  }} \\
\midrule
\multirow{2}{*}{$1$M} &  Regent & $4.72 $ & $ 0.17 $ & $4.55$ & $0.14 $ & $14.21 $\\ [0.4em]        
 & CUDA-C & $3.78 $ & $ 0.15 $ & $3.63$ & $0.13$ & $12.92 $\\ [0.2em]   
       \midrule        
\multirow{2}{*}{$32$M}&  Regent & $4.84$ & $ 0.41 $ & $4.43$ & $0.29$ & $28.53 $ \\ [0.4em]
 & CUDA-C & $4.76 $ & $ 0.68 $ & $4.08$ & $0.35$ & $34.97 $ \\ [0.2em]  \bottomrule
\end{tabular}
\end{center}
\caption{{A comparison of scheduler statistics on coarse and fine point distributions. }}
\label{naive-performance-metrics-warps}
\end{table}
Table \ref{naive-performance-metrics-warps} shows the scheduler statistics for the coarse and fine levels of point distributions. On the coarse distribution, we observe that the {\tt q\_derivatives} kernel has nearly identical {\it eligible} and {\it issued} warps per scheduler for {\tt Regent} and CUDA-C. As a result, the SM utilisation of this kernel is also similar (as shown in Table \ref{naive-performance-metrics}). However, on the fine distribution, the {\it eligible} and {\it issued} warps per scheduler in Regent are lower. This explains the lower SM utilisation of the {\tt q\_derivatives} kernel on the fine point distribution. Similarly, in the {\tt flux\_residual} kernel, the {\it eligible} and {\it issued} warps are observed to be lower than CUDA-C, resulting in poorer utilisation of SM resources in the Regent code. \\ \\ 
\begin{table}[htbp]
\centering
\begin{center}
\begin{tabular}{cccc}
\toprule
\multicolumn{1}{c}{Points} &
\multicolumn{1}{c}{Code} & 
\multicolumn{2}{c}{Warp stalls (in cycles) due to} \\
\cmidrule(lr){3-4} 
&   &  long scoreboard & wait  \\ [0.2em]
        \midrule
       \multicolumn{4}{c}{{\tt flux\_residual}} \\
       \midrule
\multirow{2}{*}{$1$M} & Regent  & $ 7.72-8.50 $ & $2.78-2.79 $ \\ [0.4em]
& CUDA-C & $ 2.26-2.92 $ & $2.95-2.96 $ \\ [0.2em]
\midrule
\multirow{2}{*}{$32$M} & Regent  & $ 3.31-3.95 $ & $2.55-2.79 $\\ [0.4em]
& CUDA-C  & $ 1.29-1.89$ & $2.89-2.90$  \\ [0.2em]       
\midrule
       \multicolumn{4}{c}{{\tt q\_derivatives}} \\
       \midrule
\multirow{2}{*}{$1$M} & Regent & $ 27.01 $ & $2.51 $  \\ [0.4em]
& CUDA-C & $ 22.16 $ & $2.81$  \\ [0.2em]
\midrule
\multirow{2}{*}{$32$M} & Regent & $ 11.16 $ & $2.55$  \\ [0.4em]
& CUDA-C & $5.68$ &  $2.72$  \\ [0.2em] \bottomrule
\end{tabular}
\end{center}
\caption{{A comparison of warp state statistics on coarse and fine levels of point distributions. }}
\label{performance-metrics-warp-stalls}
\end{table}
 From Table \ref{naive-performance-metrics-warps}, we can notice that a significant number of active warps are in a stalled state. To identify the reason behind the stalls we analyse the warp state statistics. Table \ref{performance-metrics-warp-stalls} shows the two most dominant warp stall states namely, wait and long scoreboard.  Wait-type stalls occur due to execution dependencies, which are fixed latency dependencies that need to be resolved before the scheduler can issue the next instruction. For example, consider the following chain of operations. \\ \\ 
IADD r0, r1, r2; \\ \\ 
IADD r4, r0, r3; \\ \\ 
Here, the second instruction: r4 = r0 + r3, cannot be issued until the first instruction: r0 = r1 + r2, is executed. In the present meshfree solver a signficant portion of the instructions are of FP64 type, where the current instruction depends on the result of previous instructions. As a result, we have a considerable number of stalls due to wait. Similarly, stalls due to long scoreboard usually refer to scoreboard dependencies on L1TEX operations. These operations include LSU and TEX operations. Kernels with long scoreboard warp states indicate non-optimal data access patterns. From the tabulated values, we observe that {\tt Regent} has higher stalls due to long scoreboard but lower stalls due to wait than CUDA-C.  \\ \\ 
To understand why Regent has higher stalls due to long scoreboard than CUDA-C, we analyse the pipeline utilisation. Table \ref{performance-metrics-pipe-utilisation} shows the percentage utilisation of FP$64$, ALU, and LSU pipelines. The LSU pipeline is responsible for issuing load, store and atomic instructions to the L1TEX unit for global, local, and shared memory. Compared to CUDA-C, the Regent code has higher LSU pipeline utilisation. Higher LSU pipeline utilisation corresponds to more L1TEX instructions. As a result, there are more scoreboard dependencies on L1TEX operations, leading to more stalls due to long scoreboard in the Regent code.  \\ \\ 
To further analyse the high LSU utilisation of Regent, we look at the global load and store metrics. Table \ref{naive-performance-metrics-global-load-store} shows the Global Load and Store metrics for the coarse and fine levels of point distributions. The tabulated values show that the Regent code has lower number of sectors per request than the CUDA-C code. This indicates the memory access patterns are better in Regent. However, when we compare the total number of sectors for the split kernels of {\tt flux\_residual}, Regent has almost $3$ times more load sectors and $23$ times more store sectors than CUDA-C. Similar behaviour can be observed for the {\tt q\_derivatives} kernel. The significantly higher number of global sectors in Regent correlates to the higher utilisation of the LSU pipeline and, in turn higher number of warp stalls due to long scoreboard. \\ \\ 
\begin{table}[htbp]
\centering
\begin{center}
\begin{tabular}{ccccc}
\toprule
\multicolumn{1}{c}{Points} &
\multicolumn{1}{c}{Code} & 
\multicolumn{1}{c}{FP64} &
\multicolumn{1}{c}{ALU} &
\multicolumn{1}{c}{LSU} \\ [0.2em]
 \cmidrule(lr){3-5}
   & & \multicolumn{3}{c}{in percentage} \\
\midrule
       \multicolumn{5}{c}{{\tt flux\_residual} } \\
\midrule
\multirow{2}{*}{$1$M}&  Regent & $35.46-35.51$ &  $6.98-7.39$ & $6.09-6.40$ \\ [0.4em]
 & CUDA-C & $63.54-66.61$ & $9.39-9.86$ & $2.36-2.37$ \\ [0.2em] 
\midrule
\multirow{2}{*}{$32$M} & Regent & $55.91-56.01$ & $11.09-11.10$ & $9.61-9.69$ \\ [0.4em]
& CUDA-C & $73.30-76.13$ & $12.04-12.53$ & $2.75-2.77$ \\ [0.2em]       
\midrule
       \multicolumn{5}{c}{{\tt q\_derivatives} } \\
\midrule
 \multirow{2}{*}{$1$M} & Regent & $26.60$ & $6.37$ & $5.37$ \\ [0.4em]
& CUDA-C & $27.94$ & $4.59$ & $3.58$ \\ [0.2em] 
\midrule
\multirow{2}{*}{$32$M} & Regent & $53.43$ & $12.84$ & $10.80$ \\ [0.4em]
 & CUDA-C & $69.91$ & $15.07$ & $8.99$ \\ [0.2em]       
\bottomrule
\end{tabular}
\end{center}
\caption{A comparison of pipe utilisation of the streaming multiprocessor (SM). }
\label{performance-metrics-pipe-utilisation}
\end{table}
Table \ref{performance-metrics-warp-stalls} shows that the stalls due to wait for the Regent code are lower than the CUDA-C code. The {\tt{flux\_residual}} and {\tt q\_derivatives} kernels have to execute a significant number of high-latency FP$64$ instructions that are dependent on their previous instructions. Due to these fixed latency dependencies, the compiler must insert wait instructions, allowing FP$64$ instructions to finish executing before proceeding to the next instructions. Given that our code necessitates FP$64$ computations, switching to lower-latency and lower-precision instructions is not a feasible option. Despite these fixed latency dependencies existing in both languages, the Regent code has lower stalls due to wait as it spends more time waiting on memory to be fetched. As a result, the compiler inserts less wait instructions to resolve fixed latency dependencies. This explains the lower stalls due to wait in Regent. \\ \\ 
\begin{table}[htbp]
        \centering
        \begin{center}
        \scalebox{0.70}{
        \begin{tabular}{cccccc}
            \toprule
            Points &
            \multicolumn{1}{c}{Code} & 
            \multicolumn{2}{c}{Global Load} &
            \multicolumn{2}{c}{Global Store} \\ [0.2em]
            \cmidrule(lr){3-4} \cmidrule(lr){5-6} 
& & Sectors & Sectors per request & Sectors & Sectors per request\\ [0.2em]
        \midrule
         \multicolumn{6}{c}{{\tt flux\_residual}} \\
         \midrule
\multirow{2}{*}{$1$M} & Regent & $131,501,393-132,290,250$ & $9.83-9.86$ & $23,911,932-23,920,126$ & $3.83$ \\ [0.4em]
& CUDA-C & $46,271,887-47,244,238$ & $17.51-17.99$ & $1,121,308$ & $9.00$ \\ [0.2em]  
\midrule
\multirow{2}{*}{$32$M} & Regent & $2,817,108,963-2,847,567,641$ & $7.01-7.03$ & $787,515,722-787,970,821$ & $4.07$ \\ [0.4em]
& CUDA-C & $768,406,623-802,787,448$ & $9.73-9.77$ & $35,995,540$ & $9.00$ \\ [0.2em]       
        \midrule
        \multicolumn{6}{c}{{\tt q\_derivatives}} \\
        \midrule
\multirow{2}{*}{$1$M} & Regent & $105,245,787$ & $14.96$ & $2,127,572$ & $8.50$ \\ [0.4em]
& CUDA-C & $47,244,238$ & $17.51$ & $1,121,308$ & $9.00$ \\ [0.2em]  
\midrule
\multirow{2}{*}{$32$M} & Regent & $2,087,198,922$ & $9.52$ & $68,001,128$ & $8.50$ \\ [0.4em]
& CUDA-C & $802,787,448$ & $9.73$ & $35,995,540$ & $9.00$ \\ [0.2em]
\bottomrule
        \end{tabular}
        }
        \end{center}
\caption{ A comparison of global load and store metrics on the coarse and fine levels of point distributions. }
        \label{naive-performance-metrics-global-load-store}
\end{table}
In summary, due to higher LSU utilisation, Regent code has a lower SM utilisation compared to the CUDA-C code.
Finally, we present the RDP values of the optimised Regent code. Table \ref{rdp-single-node-optimised} compares the RDP of the baseline and optimised Regent codes and the CUDA-C code. The tabulated values show that the optimised Regent code is $1.378$ times slower than the optimised CUDA-C code on the finest point distribution. 
\begin{table}[H]
\centering
\scalebox{0.80}{
\begin{tabular}{ccccccc}
\toprule 
Level & Point & Regent - Baseline & Regent - Optimised & CUDA-C & \multicolumn{2}{c}{Relative Performance}\\ [0.2em]
\midrule
 \multicolumn{5}{c}{RDP $\times$ $10^{-8}$ (Lower is better)} & \multicolumn{1}{c} { Regent - Baseline} & \multicolumn{1}{c} { Regent - Optimised} \\ [0.2em]
\cline{1-4} \cline{5-7} \\ [0.1em]
$1$ & $1$M & $1.151$ & $1.350$ & $0.846$ & $1.513$ & $1.595$\\ [0.4em]
$2$ & $2$M & $0.996$ & $0.996$ & $0.658$ & $1.495$ & $1.513$\\ [0.4em]
$3$ & $4$M & $0.885$ & $0.817$ & $0.553$ & $1.513$ & $1.477$\\ [0.4em]
$4$ & $8$M & $0.849$ & $0.716$ & $0.514$ & $1.502$ & $1.392$\\ [0.4em]
$5$ & $16$M & $0.822$ & $0.671$ & $0.498$ & $1.445$ & $1.347$\\ [0.4em]
$6$ & $32$M & $0.805$ & $0.645$ & $0.487$ & $1.445$ & $1.324$\\ [0.4em]
$7$ & $64$M & $0.799$ & $0.641$ & $0.466$ & $1.457$ & $1.378$\\
\bottomrule
\end{tabular}
}
\caption{A comparison of the RDP values of the optimised Regent and CUDA-C GPU codes. }
\label{rdp-single-node-optimised}
\end{table}
\section{Performance Analysis of the Regent Solver on CPUs }
\label{sec-per-analysis-cpus}
\begin{figure}[H]
{\centering
\includegraphics[scale=0.50,trim={0 0 10mm 5mm},angle=0,clip]{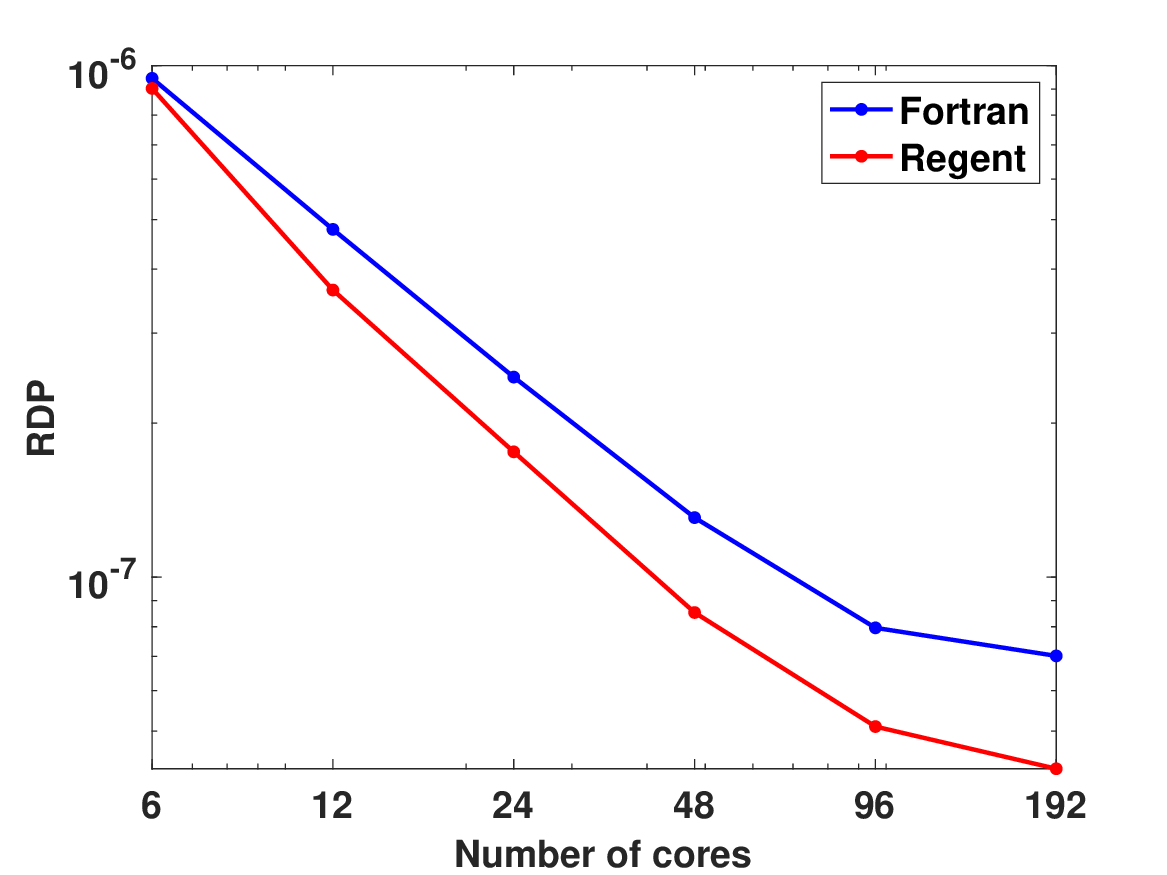}
\caption{Performance of the Regent code on a CPU node is compared with the Fortran parallel code. }
\label{cpu-rdp}
}
\end{figure}

\begin{table}[H]
\centering
\begin{tabular}{lcc}
\toprule
& CPU  \\
\midrule
Model & AMD EPYC\textsuperscript{TM} $ 9654 $ \\ [0.3em]
Cores & $192$ $\left( 2 \times 96 \right)$ \\ [0.3em]
Core Frequency & $2.40$ GHz \\ [0.3em]
Global Memory & $384$ GiB  \\ [0.3em]
$L2$ Cache & $96$ MiB \\ [0.1em]
%
\bottomrule
\end{tabular}
\caption{Hardware configuration of the CPU node. }
\label{config-cpu}
\end{table}
In this section, we assess the performance of the Regent code on CPUs by comparing its RDP values with an equivalent MPI parallel Fortran LSKUM code. For the benchmarks, numerical simulations are performed on a single point distribution around the NACA 0012 airfoil at Mach number, $M_{\infty} = 0.63$ and angle-of-attack, $AoA = 2^o$. The point distribution consists of $128$ million points. All the simulations are performed on a AMD CPU node with $192$ compute cores. Table \ref{config-cpu} shows the configuration of the CPU node. \\ \\ 
Figure \ref{cpu-rdp} shows the log-log plot of RDP values from $6$ to $192$ cores. The RDP values are computed by running the parallel codes for $1000$ fixed point iterations in eq. (\ref{state-update}). Both the codes scale well up to $96$ cores. However, on $192$ cores, the number of points per partition is insufficient to get the desired speedup. The plot shows that the Regent code performs better with increased cores as its RDP values are smaller than the Fortran code. The superior performance of Regent can be attributed to Legion's dependence analysis \cite{6468504}, which maps tasks to cores as soon as their data requirements are met. In contrast, the Fortran code employs a message-passing paradigm that uses {\tt MPI\_Barrier} to synchronise process states. In this paradigm, some processes might reach the barrier earlier than others, which results in CPU idling. Since there are no barriers in the Regent code, all the tasks will only wait for their dependencies to be resolved. This reduces the CPU idling time and explains the superior performance of the Regent code.
\section{Conclusions}
\label{sec-conclusions}
In this paper, we presented the development of a Regent based meshfree parallel solver for heterogenous HPC platforms. The meshfree solver was based on the Least Squares Kinetic Upwind Method (LSKUM) for 2D Euler equations of the inviscid fluid flow. \\ \\
The computational efficiency of the Regent solver on a single GPU was assessed by comparing it with the corresponding LSKUM solver written in CUDA-C. Benchmark simulations had shown that the Regent solver is about $1.378$ times slower than the CUDA-C solver on the finest point distribution consisting of $64$ million points. To investigate the reason behind the slowdown factor, a performance analysis of the kernels was performed. It was observed that the SM utilisation of the computationally expensive kernels in the Regent code was lower than the corresponding kernels in CUDA-C. Further analysis revealed that this was due to more long scoreboard stalls, which were caused by higher LSU utilisation in the Regent solver. \\ \\
Later, the performance of the Regent solver on CPUs was compared with an equivalent Fortran parallel LSKUM solver based on MPI. Numerical simulations on a $128$ million point distribution had shown that Regent exhibited superior performance with the increase in the number of compute cores. This was attributed to Legion's dependence analysis, which maps tasks to cores as soon as their data requirements are met. Unlike the Fortran solver, there were no barriers in the Regent solver to synchronise process states. This reduced CPU idling in Regent and thus resulted in lower RDP values. \\ \\
In summary, the performance of a single Regent LSKUM solver targeting both a GPU and CPUs is promising. Research is in progress to extend the Regent solver to three-dimensional flows.
\bibliography{ref}
\end{document}